\documentclass[prl,twocolumn,aps,showpacs]{revtex4}
\usepackage{graphicx,bm,epstopdf,latexsym,amsmath,amssymb,mathrsfs,color}

\newcommand{\pa}{\partial}

\newcommand{\ket}[1]{\left|#1\right\rangle}
\newcommand{\bra}[1]{\left\langle#1\right|}

\def\ie{\emph{i.e.},\ }

\def\ea{\emph{et~al.\ }}

\begin{document}

\title{Theory of small charge solitons in one-dimensional arrays of Josephson junctions}

\author{Stephan Rachel and Alexander Shnirman}
\affiliation{Institut f\"ur Theorie der Kondensierten Materie 
and DFG Center for Functional 
Nanostructures, Karlsruhe Institute of Technology, D-76128 Karlsruhe,
Germany}


\begin{abstract}
  We identify and investigate the new parameter regime of small charge solitons 
  in one-dimensional arrays of Josephson junctions. We obtain the dispersion relation of the
  soliton and show that it unexpectedly flattens in the outer region of the Brillouin
  zone. We demonstrate Lorentz contraction of the soliton in the middle 
  of the Brillouin zone as well as broadening of the soliton in the flat band regime. 
\end{abstract}

\maketitle

Charge solitons in one-dimensional (1D) arrays of tunnel junctions in the 
Coulomb blockade regime were introduced about twenty years 
ago~\cite{BMA89,AL91} and are being studied ever 
since (see, e.g., Ref.~\onlinecite{AGK04}).
Hermon \ea\,\cite{HBS96} studied a one-dimensional
array of Josephson junctions (JJs). It was shown that, if the grains
have a large kinetic (or geometric) inductance, the system's dynamics
are governed by the sine-Gordon model and, therefore, kink-like
topological excitations, \ie charge solitons, are the charge carriers.
Simultaneous experiments by Haviland and
Delsing\,\cite{HD96} demonstrated the Coulomb blockade 
in 1D arrays of JJs consistent with the existence of charge
solitons.  In the later experiments of Haviland's
group\,\cite{HAA00,AAH01} considerable hysteresis in the $I$-$V$
characteristic of the array was observed and attributed to a very
large kinetic inductance. The physical origin of this inductance
remained unclear. A few years later, Zorin\,\cite{Z06} pointed out
that a current biased small-capacitance JJ develops an inductive
response on top of the capacitive one. This phenomenon was called {\it
Bloch inductance}. A closely related inductive coupling between two
charge qubits was studied in Ref.\,\cite{HSMS06}. It is still not
clear if Bloch inductance could support the dynamics of charge
solitons.

In this paper, we identify the new parameter regime 
within the Coulomb blockade (insulating) phase of 
a 1D array of coupled JJs. It is
defined by the condition $\Lambda E_J > E_C > E_J$, where $E_C$ 
and $E_J$ are the charging and the Josephson energies of the junction, respectivly, and
$\Lambda$ is the bare screening length (measured in
number of junctions). In this regime we investigate the 
dynamics of charge solitons and demonstrate
two surprising features: i) flattening of the dispersion relation 
in the outer region of the Brillouin zone; ii) broadening of the soliton 
in the flat band regime in contrast to the expected and observed 
Lorenz contraction in the regime of regular dispersion relation. We believe 
these results might open the way to the explanation of the experimental 
data of Refs.~\cite{HAA00,AAH01}.

The paper is organized as follows. In order to 
shed light on the previous studies of charge solitons in
terms of the relativistic sine-Gordon equation and 
to facilitate the interpretation of our new results 
we, first, formulate the mean-field approach. Then we develop 
a many-body tight binding technique which leads to the new results.

The system considered is shown in Fig.\,\ref{Fig:Array}. The grains
are connected by JJs of capacitance $C$ (typically 1 fF) and each
grain has a capacitance $C_0$ to the ground (typically $5-20$
aF). The kinetic or geometric inductance of the grains $L_0$ is included to simplify
the mean-field treatment but it is later assumed to be vanishingly
small. We derive the following Hamiltonian:
\begin{figure}[t]
\centerline{\includegraphics[width=0.75\columnwidth]{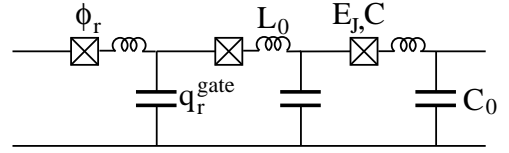}}
\caption{Josephson junction array.}
\label{Fig:Array}
\end{figure}
\begin{eqnarray}
\label{ham:array}
H &=& \sum_r \left[\frac{(2e m_r - Q_r)^2}{2C} - 
E_{\rm J}\cos\phi_r\right.\nonumber\\ &+&\left.
\frac{(Q_r-Q_{r-1})^2}{2C_0} + \frac{\Phi_r^2}{2L_0}\right]\ .
\end{eqnarray}
Here $m_r$ is the number of Cooper pairs that have tunneled
through junction number $r$. The continuous polarization charge $Q_r \equiv
\sum_{r' < r} q_{r'}^{\rm gate}$ corresponds to the integral of current 
flown into junction number $r$. The commutation relations read 
$[\Phi_r,Q_{r'}]= i\hbar \delta_{r,r'}$  and $[m_r, e^{i\phi_{r'}}] =e^{i\phi_r}\delta_{r,r'}$.

{\it Mean field approach.} In the mean field approximation we treat the
dynamical variables $Q_r$ as c-numbers, $Q_r\rightarrow\langle
Q_r(t)\rangle$.  The Hartree-like wave function can be written as a
product of single junction states,
$\Psi(\{m\})=\prod_r\Psi_{Q_r}(m_r)$. Here $\Psi_{Q}(m)$ is the (ground)
state of a single junction with Hamiltonian
\begin{equation}
\label{ham:1junction}
H_1\big(Q(t)\big) = \frac{\big(2em-Q(t)\big)^2}{2C} 
- E_J \cos\phi\ .
\end{equation}
The self-consistency condition is derived by averaging the equation 
of motion for the variables $Q_r$:
\begin{equation}
\label{eq:mean_field}
L_0 \ddot Q_r =-V_r -\frac{2Q_r - Q_{r+1}
-Q_{r-1}}{C_0}\ ,
\end{equation}
where $
V_r\equiv \langle Q_r-2e m_r\rangle/C$
is the average voltage on the junction $r$.
For static $Q_r$ and at zero temperature $V_r = \pa E_0(Q_r)/\pa Q_r$,
where $E_0(Q)$ is the lowest energy band of Hamiltonian (\ref{ham:1junction}).
Zorin\,\cite{Z06} derived an additional inductive contribution 
to the voltage on the junction:
$
V_r = \frac{\pa E_0(Q_r)}{\pa Q_r}+
L_B(Q_r) \ddot Q_r$, 
where $L_B(Q)$ is the Bloch inductance. 
Then, Eq.~(\ref{eq:mean_field}) reads 
\begin{equation}
\label{eq:SG}
L_{\rm eff}\ddot Q_r + \frac{2Q_r - Q_{r+1} - Q_{r-1}}{C_0}+
\frac{\pa E_0}{\pa Q_r} 
 = 0\ , 
\end{equation}
where $L_{\rm eff}\equiv L_0+L_B(Q_r)$. We observe that the 
inductance $L_0$ is superseded by the Bloch inductance and 
we can safely assume $L_0 = 0$.

For the case of $Q$-independent inductance $L_{\rm eff}$,
Eq.~(\ref{eq:SG}) was studied in Ref.~\cite{HBS96} (there it
was assumed that the inductance is dominated by the kinetic inductance
of the superconducting islands).  Eq.~(\ref{eq:SG}) is, then, a
discrete analog of the relativistic sine-Gordon equation and it possesses
topological solitons which describe the propagation of Cooper pairs
through the array. As usual in relativistic physics, a soliton is subject 
to the Lorentz contraction, i.e., its length reduces as its velocity grows 
(see Ref.~\cite{HBS96} and references therein).

Investigation of the case of $Q$-dependent inductance (the Bloch
inductance is a rapidly varying function of $Q$ in the regime $E_C \ge
E_J$) is still pending.  We just note
here that one could expect\,\cite{HSMS06} the effective Lagrangian of
a $Q$-biased Josephson junction to have the form $\mathcal{L}=
(1/2)L_B(Q){\dot Q}^2-E_0(Q)$. Then the voltage on the junction $r$ would
be given by $V_r = \frac{\pa E_0}{\pa Q_r}+L_B(Q_r)\ddot Q_r +
\frac{1}{2}\frac{\pa L_B}{\pa Q_r}{\dot Q_r}^2$. Thus,
Eq.~(\ref{eq:SG}) might need to be further modified.  In this paper
we do not pursue further the mean field analysis but rather
concentrate on an alternative approach of tight binding treatment of
various charge configurations.

{\it Charge configurations.}
For $L_0 \rightarrow 0$ the polarization charges $Q_r$ are enslaved to the 
discrete charges $m_r$ (the charge that have tunneled through junction $r$). 
If the charge configuration $\{m_r\}$ is given, then the polarization 
charges $\{Q_r\}$ are found from
$\frac{Q_r - 2e m_r}{C} + \frac{2Q_r - Q_{r+1}-Q_{r-1}}{C_0}=0$.
Equivalently one can consider island 
charges $n_r = m_r - m_{r+1}$ and
obtain the charging energy of the array (see, e.g., Ref.~\cite{FS91})
\begin{equation}
H_C = \frac{1}{2}\sum_{r,r'} U(r-r')\, n_r n_{r'}\ .
\end{equation}
Here
\begin{equation}\label{U-exact}
U(r) =2 E_C \int\limits_{-\pi}^{\pi} \frac{dk}{2\pi} \,
\frac{e^{ikr}}{\Lambda^{-2} - 2(\cos k -1)} \ ,
\end{equation}
where $\Lambda \equiv \sqrt{C/C_0}$ is the screening length and
$E_C\equiv (2e)^2/(2C)$ is the charging energy of a single
junction. The Josephson term in the Hamiltonian connects the charge
configurations which differ by one Cooper pair being transported
through one junction. For $\Lambda \gg 1$, the charging energy reads
$U(r)\approx
\Lambda E_C \exp{(-|r|/\Lambda)}$.

{\it Charge states nomenclature.} We consider the sector of the
Hilbert space with exactly one extra Cooper pair in the array, \ie
$\sum_r n_r=1$. The simplest representative of this sector is the
state in which the extra Cooper pair resides on island $R$ and all
other islands are neutral.  We denote this state $\ket{\ldots
  0\,0\,1_R\,0\,0\ldots}\equiv\ket{R}$. The charging energy of
$\ket{R}$ is given by $\frac{1}{2}U(0)\equiv E_0\approx \Lambda
E_C/2$. This is a rather high energy, in case of $C_0\to 0$ it is in fact
infinite (proportional to the system size\,\cite{FVB08}), and this is
approximately the energy one has to invest in order to insert the
Cooper pair into the array.  There exists, however, other charge
configurations in the single Cooper pair sector, \ie the ones with
charge--anti-charge pairs induced in the vicinity of the first Cooper
pair. The first example is the configuration $\ket{\ldots
  0\,0\,1\,-1_R\,1\,0\,0\ldots}\equiv \ket{R;1,1}$, where charge $-1$ resides 
  on island $R$ while charges $+1$ reside on the neighboring islands $R-1$ and $R+1$.
Its charging
energy is given by $E_0 + E_{1,1}$, where $E_{1,1}\equiv U(0)-2U(1)+U(2)\approx
E_C/\Lambda$. As long as $\Lambda \gg 1$ the additional energy
cost as compared to the state $\ket{R}$ is much smaller than
$E_0$. The next configurations are those of a total width $w_c=4$
($w_c$ being the number of neighboring islands involved in the
configuration), $\ket{\ldots
  0\,1\,-1_R\,0\,1\,0\ldots}\equiv\ket{R;1,2}$ and $\ket{\ldots
  0\,1\,0\,-1_R\,1\,0\ldots}\equiv\ket{R;2,1}$ with the charging
energy $E_0 + E_{1,2}$, where  $E_{1,2} \equiv U(0)-U(1)-U(2)+U(3) \approx 2E_C/\Lambda$. Thus we conclude that the regime of
dominating charging energy $E_C>E_J$ splits into two:

a) {\it Strong Coulomb blockade regime:} $E_C > \Lambda E_J$.  In this
case the charging energy difference, $\sim O(1)E_C/\Lambda$, between
the charge configurations with charge--anti-charge pairs and the basic
one $\ket{\ldots 0\,0\,1_R\,0\,0\ldots}$ is higher than the tunneling
energy $E_J$. Thus, the charge configurations of higher energy play
little role.  The basic charge configurations form a trivial tight
binding band with dispersion $E(k) = -E_J \cos{k}$. It is this
regime which was analyzed in 2D in Ref.~\cite{SEA09}.

b) {\it Small solitons regime:} $\Lambda E_J > E_C > E_J$. In this
case several charge configurations hybridize with the basic one and
small solitons are formed.  In what follows, we investigate this
regime and we develop a tight binding approach which allows us to
treat this case numerically.  A similar approach for polarons was
developed in Ref.~\cite{BTB99}.

\begin{figure}[t!]
\centering
\includegraphics[scale=0.2]{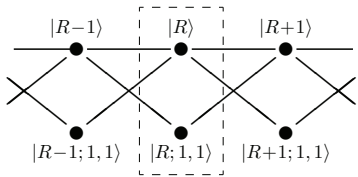}
\caption{Effective lattice and effective unit cell for the
  configurations $\ket{R}$ and $\ket{R;1,1}$. The dashed box marks a
  primitive unit cell. The lines denote allowed tunneling between
  the configurations.}
\label{fig:2x2}
\end{figure}

\begin{figure}[b!]
 \centering
\includegraphics[scale=0.6]{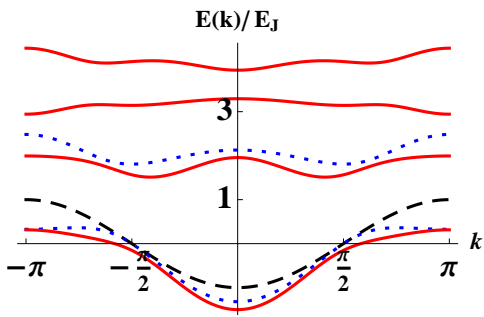}
\caption{(Color online) Dispersion relation for the one-state-approximation,
  \ie a single Cooper pair (black dashed band), for the
  two-state-approximation (blue dotted  bands), and for the
  four-state-approximation (red solid bands) as described in the text.  We
  chose $E_C = 20 E_J$, $\Lambda=10$.}
\label{fig:124bands}
\end{figure}

To illustrate our approach we start by accounting only for two
configurations, $\ket{R}$ and $\ket{R;1,1}$. In Fig.\,\ref{fig:2x2}
the structure of possible transitions between these states by
tunneling of a single Cooper pair is shown.  We observe that a tight
binding situation arises again with two states per primitive unit
cell.
Instead of the $\cos{k}$-dispersion, we obtain the following $2\times
2$ matrix
\begin{equation}
 H^{(2)}_k = -E_J\left( \begin{array}{cc} \cos{k}&\cos{k}\\[5pt]
 \cos{k}&0\end{array}\right)+
\left( \begin{array}{cc} E_0&0\\[5pt]
 0&E_0+E_{1,1}\end{array}\right)\ ,
\end{equation}
where the second matrix accounts for the charging
energies of the states $\ket{R}$ and $\ket{R;1,1}$. 
In what follows we omit the common energy $E_0$ for all states. 
Diagonalizing $H^{(2)}_k$ yields two bands as shown
in Fig.\,\ref{fig:124bands} (blue dotted curves). 
Next we add the charge states 
$\ket{\ldots 0\,1\,-1_R\,0\,1\,0\ldots}\equiv\ket{R;1,2}$ and
$\ket{\ldots 0\,1\,0\,-1_R\,1\,0\ldots}\equiv\ket{R;2,1}$. 
We find the $4\times 4$ tight binding
matrix $H_k^{(4)}=$
\begin{equation}
\small
-E_J\!\left(\begin{array}{cccc} 
\cos{k}&\cos{k}&\frac{1}{2}\exp{(-2ik)}&\frac{1}{2}\exp{(2ik)}\\[5pt]
\cos{k}&-\frac{E_{1,1}}{E_J}&\frac{1}{2}&\frac{1}{2}\\[5pt]
\frac{1}{2}\exp{(2ik)}&\frac{1}{2}&-\frac{E_{1,2}}{E_J}&\frac{1}{2}
\exp{(ik)}\\[5pt]
\frac{1}{2}\exp{(-2ik)}&\frac{1}{2}&\frac{1}{2}\exp{(-ik)}&-\frac{E_{1,2}}
{E_J}
\end{array}\right)\ .
\nonumber
\end{equation}
In Fig.\,\ref{fig:124bands}, the
single particle band, the two bands of $H_k^{(2)}$, and the four bands
of $H_k^{(4)}$ are shown for $E_C=20 E_J$ and $\Lambda=10$. Here we
are clearly in the strong Coulomb blockade regime and inclusion of the 
extra states only slightly modifies the lowest energy band. 

\begin{figure}[b!]
\centering
\includegraphics[scale=0.95]{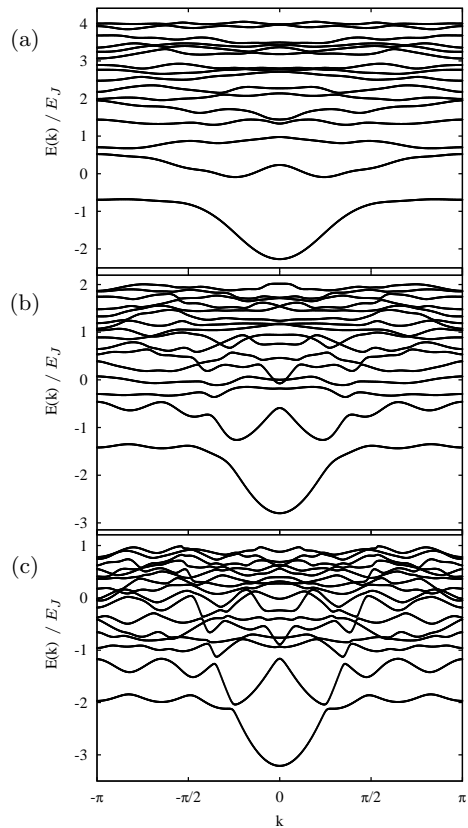}
\caption{Band structure for $w_c=7$ with parameters $\Lambda=10$ and 
a) $E_C=10 E_J$, b) $E_C=5 E_J$, c) $E_C=2.5 E_J$. 
For clarity, only the 16 lower bands are shown.}
 \label{fig:32bands}
\end{figure}

The idea is now to approach the intermediate regime $\Lambda E_J > E_C
> E_J$ by extending the number of charge configurations. Here we went
up to the total width of the charge configurations $w_c=7$ resulting
in a $32\times 32$ tight binding matrix. We investigate three regimes 
$E_C=10E_J$, $5E_J$, and $2.5E_J$ ($\Lambda=10$). 
The resulting spectra are shown in
Fig.\,\ref{fig:32bands}. 
While in the strong Coulomb blockade regime $E_C > \Lambda E_J$ (see
Fig.\,\ref{fig:124bands}) the lowest band is very close to the
$-cos{(k)}$ dispersion of a free particle, the shape of the lowest
band in the regime of small solitons $E_C \le \Lambda E_J$ (see
Fig.\,\ref{fig:32bands}) changes considerably. For $E_C/E_J=10$ which
is the upper boundary of the "small soliton" regime the lower band
still has the cosine-shape for $|k|<\pi/2$. For larger values of
$|k|$, however, the band becomes very flat which corresponds to zero
group velocity or, equivalently, to infinite mass.  For smaller ratios
$E_C/E_J$, we find that the region in the center of the
Brioullin-zone, which is cosine-like or parabolic, becomes smaller
($|k|<\pi/4$ for $E_C/E_J=2.5$). The remaining flat region shows a
weak oscillatory behavior. We cannot exclude that it is due to an 
insufficient number of charge configurations 
included. Indeed, while for $E_C/E_J = 10$ the numerical convergence for the lowest band 
is good, it somewhat deteriorates for smaller values of $E_C$. For $E_C=2.5E_J$ the first and second bands
approach each other at $|k|\approx\pi/4$. This could give rise to
Landau-Zener transitions for an accelerated soliton.

\begin{figure}[t!]
\centering
\includegraphics[scale=0.9]{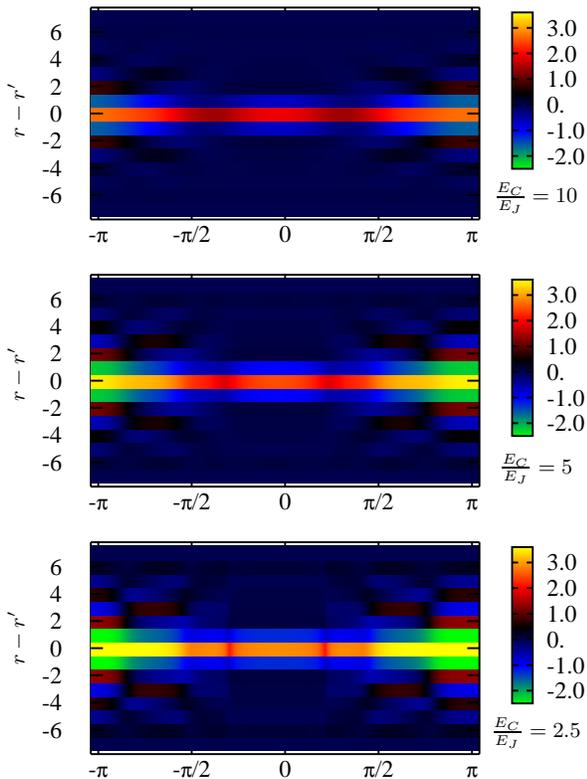}
\caption{(Color online) Charge-charge correlator $\bra{\psi_k} n_r n_{r'}
  \ket{\psi_k}$ for $E_C=10 E_J$, $5E_J$, and $2.5E_J$. In all plots
  $\Lambda = 10$.}
\label{fig:chargedist}
\end{figure}

\begin{figure}[ht!]
\centering
\includegraphics[scale=0.8]{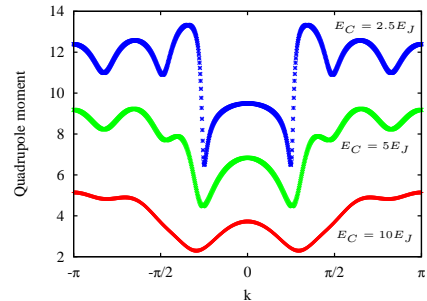}
\caption{Quadrupole moment of the charge-charge correlator $\bra{\psi_k}
 n_r n_{r'}\ket{\psi_k}$ for $E_C=10 E_J$, $5E_J$, 
and $2.5E_J$. For
 $|k| < \pi/4$ the Lorentz contraction can be observed.}
 \label{fig:quadrupole}
\end{figure}

{\it Soliton shape.}
We investigate the charge smearing in the regime of small
solitons. For that purpose, we consider the charge-charge correlation
function $F(k,r-r')=\bra{\psi_k} n_{r} n_{r'}\ket{\psi_k}$, where
$\ket{\psi_k}=\sum_{R,j} \alpha_j(k)\ket{R,j}e^{ikR}$ is the Bloch
wave function of the soliton (lowest band). Here $\ket{R,j}$ denotes
the $j$-th charge configuration centered at the island $R$, 
e.g., $\ket{R}$, $\ket{R;1,1}$, $\ket{R;1,2}$ etc.. 
We obtain $F(k,r-r')=\sum_j
\, |\alpha_j(k)|^2 \,\mathcal{C}_j(r-r')$, where $
\mathcal{C}_j(r)\equiv \sum_{r'} n^{R,j}_{r'} n^{R,j}_{r'+r}$ (this quantity is independent of R).  
The quantity $n^{R,j}_r$ is the number of charges on island $r$
for the charge configuration $\ket{R,j}$.  Note that the correlation function
is normalized, \ie $\sum_{r'}\langle n_r n_{r'}\rangle=1$
if we choose the normalization of the Bloch wave functions such that
$\sum_j |\alpha_j|^2 =1$.  In Fig.\,\ref{fig:chargedist} we plot the
charge-charge correlation function in the whole Brillouin zone.  
We observe extended structure appearing in the flat band regions. To
characterize the width of the charge distribution we plot in
Fig.\,\ref{fig:quadrupole} the quadrupole moment
$\mathcal{Q}(k)=\sum_{r} {r}^2 F(k,r)$.  For
small values of $k$ we observe the Lorentz contraction, as predicted
by the sine-Gordon model. In the region of flat dispersion (infinite
mass) the soliton becomes much wider. A question arises whether a model 
of sine-Gordon type could explain this phenomenon.

{\it Discussion.} In this paper, we have identified the regime of small charge solitons
and investigated numerically their properties. One of the
characteristic features is the flattening of the dispersion relation
in the outer region of the Brillouin zone and simultaneous broadening of the
soliton. 

Our study was performed for infinite arrays with no disorder (offset charges).
In the limit $\Lambda \gg 1$, both, the array borders and the offset charges 
create smooth variations  of the potential energy of a Cooper pair (wells or barriers). 
The amplitude of these variations $\sim O(1)\Lambda E_C$ is, however, very large. 
The propagation of charge will thus crucially depend on the dispersion relations 
obtained in this paper as well as on the dissipation in the system. Further 
studies of these issues are necessary.

We acknowledge numerous discussions with A. Ustinov, R. Sch\"afer,
H. Rotzinger, and B. Malomed as well as the participants of the SCOPE
2009 meeting in Karlsruhe. We thank I. Martin for pointing
Ref.~\cite{BTB99} to us.


\end{document}